\documentclass[aps,prb,twocolumn,groupedaddress,showpacs]{revtex4}
\usepackage{graphicx,amsmath,subfigure}

\newcommand{\keybox}[1]{
\protect\raisebox{-.4ex}[0pt][0pt]{
\protect\includegraphics[width=1em]{#1}}}

\begin{document}

\title{Stability of strained heteroepitaxial systems in $(1+1)$
dimensions}

\author{Pierre Thibault}
\email[Present address: Laboratory of Atomic and Solid State Physics, Cornell
University, Ithaca NY 14825, U.S.A. ]{e-mail: thibault@physics.cornell.edu}
\affiliation{D\'epartement de physique et Regroupement qu\'eb\'ecois sur les
mat\'eriaux de pointe (RQMP) \\ Universit\'e de Montr\'eal, Case Postale
6128, Succursale centre-ville, Montr\'eal, Qu\'ebec, Canada, H3C 3J7}

\author{Laurent J. Lewis}
\email{laurent.lewis@umontreal.ca}
\affiliation{D\'epartement de physique and Regroupement qu\'eb\'ecois sur les
mat\'eriaux de pointe (RQMP) \\ Universit\'e de Montr\'eal, Case Postale
6128, Succursale centre-ville, Montr\'eal, Qu\'ebec, Canada, H3C 3J7}

\date{\today}

\begin{abstract}
We present a simple analytical model for the determination of the stable
phases of strained heteroepitaxial systems in $(1+1)$ dimensions. In order
for this model to be consistent with a subsequent dynamic treatment, all
expressions are adjusted to an atomistic Lennard-Jones system. Good agreement
is obtained when the total energy is assumed to consist of two contributions:
the surface energy and the elastic energy. As a result, we determine the
stable phases as a function of the main ``control parameters'' (binding
energies, coverage and lattice mismatch). We find that there exists no set of
parameters leading to an array of islands as a stable configuration. We
however show that a slight modification of the model can lead to the
formation of stable arrays of islands.
\end{abstract}


\maketitle

\section{Introduction \label{sec:intro}}

It appears today that self-assembly is not only one of the most elegant
avenues for the production of devices based on quantum dots (QD), but also
one of the most promising. Basic understanding of the physics of the
formation of arrays of islands should ultimately lead to the realization of
such exciting concepts as spintronics\cite{datta1990,loss1998} and quantum dot
cellular automata.\cite{lent1993} Driven by these possible developments,
considerable effort has been devoted to understanding and predicting the
conditions necessary for ensuring the stability of arrays of islands. Simple
arguments based on the scaling of the energy as a function of the volume of
the islands\cite{combe2001} indicate that any system naturally undergoes
ripening and, therefore, the only relevant observation is the (very long)
time scale needed for the system to reach equilibrium.\cite{wang1999} It is
becoming clear, however, that a realistic energy function can lead to arrays
of islands as equilibrium configurations,\cite{shchukin1995, daruka1997,
zhang2000, combe2001, prieto2003} as can also be deduced from experiment (see,
e.g., Ref.\ \onlinecite{shchukin1999} for a review).

The dynamics of formation of arrays of islands has to some extent been
investigated using atomistic models,\cite{barabasi1997, koduvely1999,
khor2000, meixner2001} but is not yet fully understood. For instance, the
importance of nucleation in the early stage of array formation is still
unclear.\cite{tromp2000} The long-term goal of our work is to address such
questions and to provide a coherent picture of island formation in
heteroepitaxial systems, duly taking into account changes in the energy
landscape arising from the lattice mismatch between the two components of the
system. We aim to achieve this using a kinetic Monte Carlo (KMC) model,
whereby the particles evolve according to the relative probabilities for
hopping from one site to a neighbouring site on a fixed lattice. The main
difficulty of this approach, which we are still in the process of developing,
resides in properly modulating the energy barriers to account for elastic
contributions generated by the lattice mismatch between the two species. We
note that this problem could in principle be approached using molecular
dynamics (MD). However, the timescales involved in the present problem are
such that MD calculations are not feasible at this time.

For lack of an accurate model for the kinetics of island formation, there is
definite interest in the investigation of the equilibrium properties of
heteroepitaxial systems, i.e., the surface morphology as determined by the
various parameters that control the physics of the systems (lattice mismatch,
coverage, binding energies, etc.). Our objective here is to present and
discuss a continuous model suited for this purpose. This will enable us, in
particular, to identify the regions of parameter space where the formation of
stable, coherent arrays of island are favored. We develop an analytical
expression for the zero-temperature total energy of a strained array of
islands. To ensure consistency, an important constraint on this model is that
it should be expressed in terms of quantities that can be ``exported'' to a
subsequent KMC model which will allow the dynamics to be investigated. We do
this by assuming Lennard-Jones interactions between the two types of atoms
and considering a triangular ``$(1+1)$-dimensional'' geometry. This defines a
reference system (which we will call the {\em LJ system}) on which both the
static and the dynamic models should be mapped. Note that we use the
term ``$(1+1)$-dimensional'', rather than ``2-dimensional'', to
indicate that one of the spatial dimensions is height (the $z$ component),
not to be confused with the usual two-dimensional case where atoms move in
the $xy$ plane.

We thus obtain an expression for the difference in energy $\Delta E$ between
a system with a flat layer of adsorbed atoms and a system where islands have
formed. For a given set of control parameters, this quantity is minimized
with respect to the size of the islands, the distance between the islands,
and the thickness of the wetting layer, so as to determine the equilibrium
state of the system. We find that good agreement between the LJ system and
the continuous model can be achieved by considering only two contributions in
$\Delta E$: the surface energy and the elastic energy, the latter arising
from both relaxation and substrate-mediated island-island interactions. This
general approach is in many respects similar to that proposed by Combe et
al.\cite{combe2001} (CJB). Apart from the fact that our model allows the
lattice misfit and the energy parameters to vary, the main differences lie in
the details of the method, as discussed below.

An important conclusion of our work is that {\em no single set of control
parameters} leads to an array of islands as a stable configuration. This is
in contradiction with the results of CJB;\cite{combe2001} the discrepancy
arises from the arbitrariness in the choice of the parameter ($z_0$) which
represents the characteristic length for the decay of the adsorption energy
above the surface. If we relax the constraint on the choice of this parameter
(i.e., consistency with the LJ model is not imposed), then stable arrays of
islands are found, along with a ``cracks'' phase, made of flat islands that
touch at their base.

\section{The Model \label{sec:model}}

\subsection{The LJ system \label{subsec:lj}}

As mentioned above, the reference system consists of atoms occupying the
sites of a $(1+1)$-dimensional triangular lattice and interacting via the
Lennard-Jones potential:
   \begin{equation}
   U(r) = 4 \epsilon \left[ \left( \frac{\sigma}{r} \right)^{12} - \left(
   \frac{\sigma}{r} \right)^{6} \right].
   \label{eq:lj}
   \end{equation}
The atoms are of two types: substrate ($S$) and adsorbate ($A$). Thus, there
are three different types of interactions and six different Lennard-Jones
parameters that define the total energy: $\{\sigma_{SS}, \sigma_{AA},
\sigma_{SA}\}$ and $\{\epsilon_{SS}, \epsilon_{AA}, \epsilon_{SA}\}$. Since
one of the $\sigma$'s and one of the $\epsilon$'s fix the length scale and
energy scale, respectively, there are only four free parameters. This number
can be reduced to three by applying the Lorentz-Berthelot combination rule:
   \begin{equation}
   \sigma_{SA} = \frac{1}{2} (\sigma_{SS} + \sigma_{AA}).
   \label{eq:combin}
   \end{equation}
Hence, there is a single degree of freedom as far as length scales are
concerned, which can be expressed in terms of the lattice mismatch $\alpha$,
defined as
   \begin{equation}
   \alpha = \frac{\sigma_{AA} - \sigma_{SS}}{\sigma_{SS}}.
   \label{eq:alpha}
   \end{equation}
The other two degrees of freedom are the binding energies between adsorbate
atoms, $\epsilon_{AA}$, and between adsorbate and substrate atoms,
$\epsilon_{SA}$. These parameters are independant and $\epsilon_{SA} >
\epsilon_{AA}$ is the wetting condition.

In practical calculations, the LJ interaction must be cutoff at some distance
$r_c$. The choice of $r_c$ affects slightly the physical properties of the
system, notably the equilibrium interatomic distance $r_{\text{eq}}$, the
cohesive energy $u_0$, and the elastic constants. (See Section
\ref{subsubsec:elast} for details on the calculation of the elastic
constants). Table \ref{tab:1} presents the dependence of these important
quantities on the cutoff radius.

\begin{table}[ht!]
\centering
\begin{tabular}{c|c|c|c}
 $r_c$    & $r_{\text{eq}}$ ($\sigma$) & $u_0$ ($\epsilon$) & $\mu = \lambda$
($\epsilon$) \\ 
\hline
1.000 (1) & 1.1225              & -3.000             &  31.18             \\ 
1.732 (2) & 1.1159              & -3.222             &  33.48             \\ 
2.000 (3) & 1.1132              & -3.319             &  34.49             \\ 
2.646 (4) & 1.1122              & -3.356             &  34.87             \\ 
3.000 (5) & 1.1119              & -3.364             &  34.96             \\ 
$\infty$  & 1.1115              & -3.382             &  35.15             \\ 
\end{tabular}
\caption{Computed values of some important quantities as a function of the
cutoff distance $r_c$ (and number of nearest-neighbor shells):
$r_{\text{eq}}$ is the equilibrium interatomic spacing in units of $\sigma$,
$u_0$ is the cohesive energy in units of $\epsilon$, and $\mu$ and $\lambda$
are the two Lam\'e parameters (both are equal in this geometry).}
\label{tab:1}
\end{table}

\subsection{The Continuous Model \label{subsec:contin}}

\begin{figure}
\includegraphics[width=8.5cm]{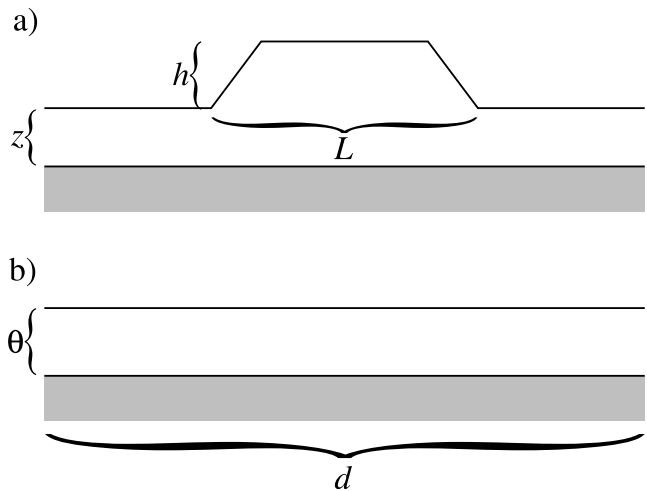}
\caption{Geometry of the system (a) with islands and (b) without islands. The
shaded region is the substrate and the white region is the adsorbate. Islands
are assumed to have the shape of an isosceles trapezoid, with contact angle
$\frac{\pi}{3}$.}
\label{fig:geometry}
\end{figure}

As discussed in the Introduction, our purpose is to evaluate the energy
difference between a system in which the adsorbate atoms form islands and one
in which they form a uniform layer on top of the substrate:
   \begin{equation}
   \Delta E = E_{\text{island}} - E_{\text{layer}}.
   \label{eq:deltae}
   \end{equation}
This energy difference can be decomposed into surface and elastic
contributions,
   \begin{equation}
   \Delta E = \Delta E_{\text{surface}} + \Delta E_{\text{elastic}},
   \label{eq:deltae1}
   \end{equation}
and is a function of the following parameters:
   \begin{itemize}
   \item $\alpha$, the lattice mismatch;
   \item $\epsilon_{SS}$, the binding energy between two atoms of the
         substrate;
   \item $\epsilon_{AA}$, the binding energy between two atoms of the
         adsorbate;
   \item $\epsilon_{SA}$, the binding energy between an atom of the
         substrate and an atom of the adsorbate;
   \item $\theta$, the coverage, expressed in monolayers (ML);
   \item $h$, the height of the islands, expressed in ML;
   \item $L$, the width of the islands at their base;
   \item $z$, the thickness of the wetting layer, expressed in ML;
   \item $d$, the distance between the centers of two neighboring islands.
   \end{itemize}
As mentioned earlier, one of the binding energies, say $\epsilon_{SS}$, fixes
the energy scale. The last five parameters describe the geometry of the
system (cf. Fig.\ \ref{fig:geometry}); they are integers but we will assume
that they are real in order to facilitate the calculations. Since $h$,
$\theta$ and $z$ are expressed in ML, the actual height is obtained by
multiplying by the thickness of one monolayer, $\frac{\sqrt{3}}{2} r_{eq}$.

The conservation of atoms (or volume) between the two configurations imposes
a constraint on the geometric parameters:
   \begin{equation}
   \theta d = zd + h ( L - h/2 ).
   \label{eq:constraint}
   \end{equation}
Since $\alpha$, $\epsilon_{SS}$, $\epsilon_{SA}$, $\epsilon_{AA}$ and
$\theta$ are assumed to be known from experiment for a given material, i.e.,
they can be considered as control parameters, the energy need only be
minimised with respect to the three parameters $L$, $h$ and $z$, $d$ being
determined by the constraint \eqref{eq:constraint}.

\begin{figure}[!ht]
\includegraphics[width=6cm]{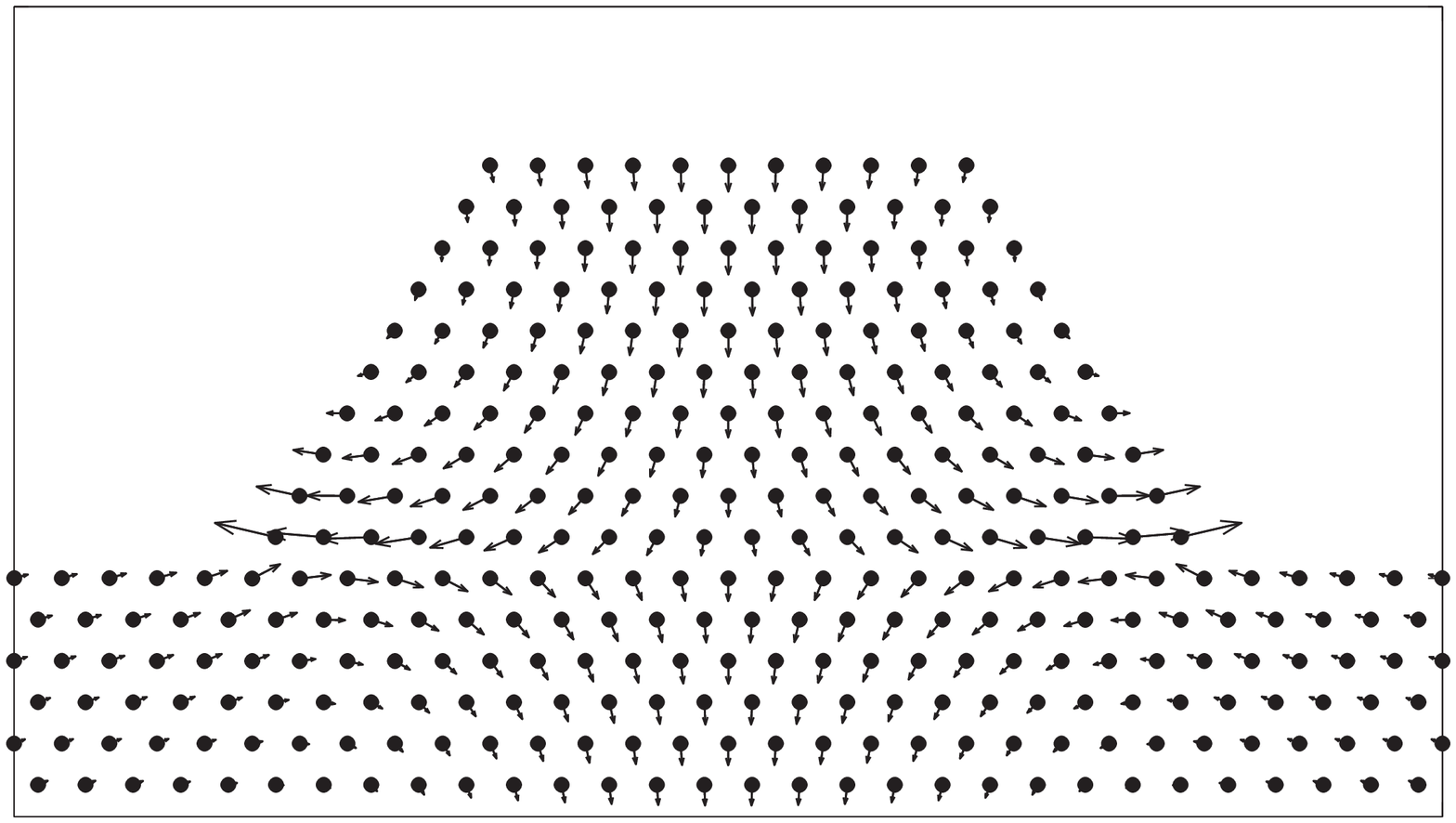}
\raisebox{3cm}{\makebox[0pt][r]{a) \hspace{5.5cm}}}
\includegraphics[width=6cm]{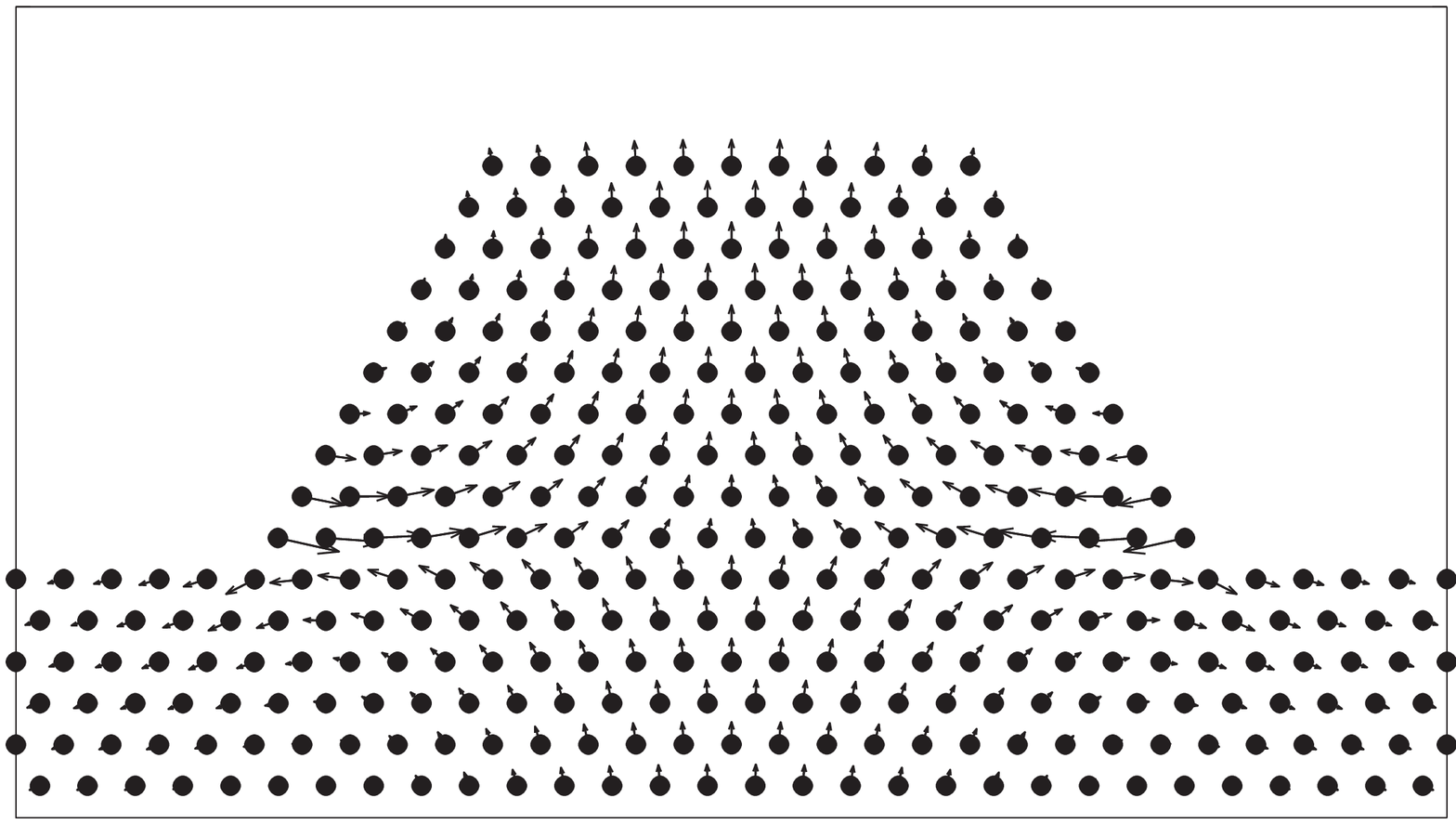}
\raisebox{3cm}{\makebox[0pt][r]{b) \hspace{5.5cm}}}
\caption{Displacements of the atoms of a relaxed island ($L=20, h=10$).
(a) tensile ($\alpha = -1\%$); (b) compressive ($\alpha = +1\%$). The
lengths of the arrows are about 30 times the actual atomic displacements.}
\label{fig:displacement}
\end{figure}

In the next two subsections we develop expressions for the two energy
contributions in Eq.\ \eqref{eq:deltae1} as a function of the various
parameters of the problem. They are derived largely from theoretical
considerations but, in some cases, {\em ad hoc} terms are introduced in order
to ensure that the model is consistent with the LJ system; in these cases, we
proceed as follows: first, we generate a set of configurations with some
number of adsorbate atoms placed in the shape of a trapezoid. Then, with
suitably chosen parameters ($\epsilon$ and $\sigma$), we let the whole system
(including a substrate large enough to make finite size effects negligible)
relax to minimum energy. Periodic boundary conditions are used along the $x$
direction. Figure~\ref{fig:displacement} illustrates the atomic displacement
of the atoms of an island as a result of relaxation.

\subsubsection{The Surface Energy \label{subsubsec:surf}}

\begin{figure}[b]
\includegraphics[width=8.5cm]{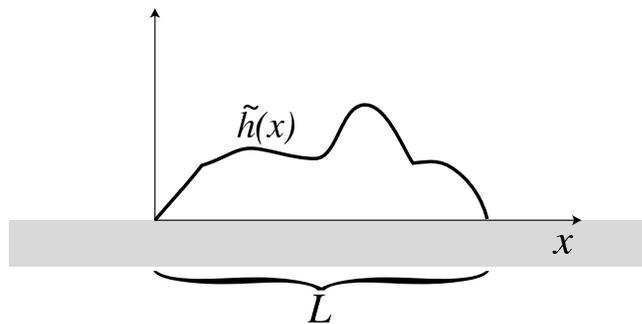}
\caption{Island of shape given by $\tilde h (x)$.}
\label{fig:island}
\end{figure}

We first determine the adsorption energy of an island of type $S$ (i.e.,
``substrate on substrate'') whose actual height is $h(x)$, or $\tilde
h(x)$ when expressed in ML (see Fig. \ref{fig:island}). This energy is
the sum of bulk and surface contributions:
   \begin{equation}
   E^{\text{island}}_S = V u_0 \epsilon_{SS} + (\tilde L - L)\gamma_{SS},
   \label{eq:enadsorps}
   \end{equation}
where $V$ is the volume of the island, $u_0 \epsilon_{SS}$ the cohesive
energy per atom, $(\tilde L - L)$ the increase of surface due to the island,
and $\gamma_{SS}$ the surface energy density. Using $\tilde L = \int
\sqrt{dx^2 + dz^2} = \int \sqrt{1 + h'^2(x)} dx$, $E^{\text{island}}_S$ can
readily be rewritten in terms of $\tilde h(x)$:
   \begin{equation}
   E^{\text{island}}_S \! = \! \int_0^L \!\! \left[ \tilde h u_0
   \epsilon_{SS} + \! \left( \! \sqrt{ \frac{3}{4} \tilde h'^2 + 1} - \! 1
   \! \right) \! \gamma_{SS} \right] \! dx,
   \label{eq:enadsorps1}
   \end{equation}
where the factor $\frac{3}{4}$ comes from the substitution of $h(x)$ by
$\frac{\sqrt{3}}{2} \tilde h(x)$.

The surface energy density $\gamma_{SS}$ is proportional to $\epsilon_{SS}$;
we can write
   \begin{eqnarray}
   \gamma_{SS} &=& C \epsilon_{SS} \nonumber \\
   \gamma_{AA} &=& C \epsilon_{AA} \label{eq:surfendens} \\
   \gamma_{SA} &=& C (\epsilon_{SS} + \epsilon_{AA} - 2
   \epsilon_{SA}), \nonumber
   \end{eqnarray}
where $C$ is a constant that depends on the number of neighbors taken into
account in the model.

If the island is of type $A$ (i.e., adsorbate on substrate), now, the adsorption
energy is easily obtained from Eq.\ \eqref{eq:enadsorps1} by substituting
$\epsilon_{AA}$ and $\gamma_{AA}$ for $\epsilon_{SS}$ and $\gamma_{SS}$ and
adding a term describing the interaction between the island and the
substrate. If we assume nearest-neighbor interactions, this term involves
only the atoms at the interface between the island and the substrate. In a
more general situation, the adsorption energy for an atom at position $z$
above the surface (in ML) can be written
   \begin{equation}
   E_{\text{ad}}(z) = E^0_{\text{ad}} + (\epsilon_{AA}-\epsilon_{SA})\eta(z),
   \label{eq:adsorption}
   \end{equation}
where $E^0_{\text{ad}}$ is the ``generic'' adsorption energy for the case
$\epsilon_{SA} = \epsilon_{AA}$ and $\eta(z)$ is a function which decreases
with $z$, with characteristic length $z_0$. As in similar
works\cite{daruka1997, muller1997, combe2001} we assume the form:
   \begin{equation}
   \eta(z) = A e^{-z/z_0};
   \label{eq:defeta}
   \end{equation}
the parameters $A$ and $z_0$ can be determined by fitting to the total energy
of a particular atomic model (here Lennard-Jones). Using
\eqref{eq:adsorption}, we find that the adsorption energy of a vertical
column of $h$ atoms is proportional to
   \begin{equation}
   \sum_{j=1}^h e^{-j/z_0} = \frac{1-e^{-h/z_0}}{e^{1/z_0} - 1}.
   \end{equation}
Hence, we obtain:
   \begin{multline}
   E^{\text{island}}_A = \int_0^L \! \left[ \tilde h u_0
   \epsilon_{AA} +  \left(  \sqrt{ \frac{3}{4} \tilde h'^2 + 1} -  1
   \right) \gamma_{AA}  \right. \\
    \left. + B (\epsilon_{AA} - \epsilon_{SA}) \left( 1 - e^{-\tilde
   h / z_0} \right) \right] \! dx,
   \label{eq:enadsorpa}
   \end{multline}
where $B = A/(e^{1/z_0} -1)$. It can easily be shown that $B = 2C$ by
taking the limiting case of a very thick adsorbate. Overall, therefore, only
two parameters need to be fitted to the atomic model, namely $B$ and
$z_0$. This fit was carried out on systems with uniform adsorbed layers
of thickness $\theta$ [i.e., $\tilde h (x) = \theta \Rightarrow \tilde h'(x) =
0$]. In this case, the energy difference between a system with an
adsorbate of type $A$ and an adsorbate of type $S$ is
   \begin{multline}
   E^{\text{layer}}_A - E^{\text{layer}}_S = d [\theta u_0 (\epsilon_{AA}
   - \epsilon_{SS}) \\ + B(\epsilon_{AA} -
   \epsilon_{SA})(1-e^{-\theta/z_0})].
   \end{multline}
The numerical calculations yield $B=2.53$ and $z_0 = 0.39$ (see
Appendix~\ref{app:1} for details).

Assuming $\tilde h(x)$ describes a trapezoidal shape, we can finally
write the first term of \eqref{eq:deltae1} as
   \begin{multline}
   \Delta E_{\text{surface}} = 2C(\epsilon_{AA} - \epsilon_{SA}) [ d - (d-
   L + z_0)e^{-z/z_0} \\ - (L-h-z0)e^{-(h+z)/z0} - d(1-e^{-\theta/z0})] + C
   \epsilon_{AA} h.
   \label{eq:esurface}
   \end{multline}

\subsubsection{Elasticity \label{subsubsec:elast}}

It is an interesting fact that the theory of elasticity, which deals with
continuous media, can accurately describe systems as small as a few tens or
hundreds of atoms (see Ref. \onlinecite{wittmer2002} for instance). In
this work, we exploit this property to construct (at least in part) the analytical
expressions entering the second term of \eqref{eq:deltae1}. Unfortunately, we
know of no analytical solution to the elasticity differential equations for a
system with the boundary conditions illustrated in Fig. \ref{fig:geometry}a).
This difficulty can be circumvented by making some assumptions on the
force distribution caused by the island on the substrate. As argued by
Tersoff and Tromp,\cite{tersoff1993} when the island is very flat ($h
\ll L$), one can assume that there is no strain relaxation in the $z$
direction. This leads to a force distribution which is proportional to
the derivative of the height function, $f \propto h'(x)$. In more
general cases, however, we found that a (truncated) linear force density
was a better assumption. Since higher islands deform the substrate more
efficiently (for a given volume), we used the following expression:
   \begin{equation}
   f(x) = \begin{cases} kx & \text{if $\left| x \right| <
   \frac{L}{2}$} \\ 0 & \text{otherwise,} \end{cases}
   \label{eq:fdist}
   \end{equation}
where $k$ is a constant (to be determined) which depends on the lattice
mismatch, the shape of the island and the elastic modulii of both the island
and the substrate.

Following the guidelines presented in Ref.~\onlinecite{landau1995}, we
find that the Green's tensor for a semi-infinite two-dimensional plane is
   \begin{equation}
   \begin{split}
   G_{xx} &= \frac{\left( \lambda + \mu \right) x^2 - \left( \lambda + 2
   \mu \right) r^2 \log{(r)}}{2 \pi \mu \left( \lambda + \mu \right)
   r^2} \\
   G_{xz} &= \frac{\left( \lambda + \mu \right) x z - \mu r^2
   \arctan{(\frac{x}{z})}}{2 \pi \mu \left( \lambda + \mu \right) r^2} \\
   G_{zx} &= \frac{ \left( \lambda + \mu \right) x z + \mu r^2
   \arctan{(\frac{x}{z})}}{2 \pi \mu \left( \lambda + \mu \right) r^2} \\
    G_{zz} &= \frac{- \left( \lambda + \mu \right) x^2 - \left(
   \lambda + 2 \mu \right) r^2 \log{(r)}}{2 \pi \mu \left( \lambda + \mu
   \right) r^2},
   \end{split}
   \label{eq:gij2d}
   \end{equation}
where $r^2 = x^2 + z^2$. The $x$ component of the displacement field on
the surface of the substrate is found to be
   \begin{eqnarray}
   u_x(x,z \! = \! 0) &= \int_{-\infty}^{\infty} \! G_{xx} (x,z\! = \!
   0) f(x-x') dx' \nonumber \\
   &\equiv \frac{L}{2} \tilde u \big( \frac{2x}{L} \big),
   \label{eq:defux}
   \end{eqnarray}
where $\tilde u(\xi)$ is a scale-independant function:
   \begin{equation}
   \tilde u(\xi) = \kappa \left[ 2\xi + \left( 1 - \xi^2 \right)
   \log{\left| \frac{1+\xi}{1-\xi} \right|} \right],
   \label{eq:utilde}
   \end{equation}
with $\kappa = \frac{3kL}{16\pi \mu_{S}}$ and $\mu_{S}$ is one of the Lam\'e
parameters of the substrate.

Because the atoms lie on a triangular lattice and interact via a radial
potential, the two Lam\'e parameters $\mu$ and $\lambda$ must be equal, given
by
   \begin{equation}
   \lambda = \mu = \frac{\sqrt{3}}{8} \sum_j r_j^2 U''(r_j),
   \label{eq:deflame}
   \end{equation}
the sum being carried out over all lattice points within a given cutoff
radius. The Young modulus and Poisson ratio for a 2-dimensional system
are\footnote{Note the error in Ref. \onlinecite{combe2001}.}
   \begin{equation}
   \begin{split}
   E &= \frac{4 \mu (\lambda+\mu)}{\lambda+2\mu} = \frac{8}{3}\mu \\
   \nu &= \frac{\lambda}{\lambda + 2\mu} = \frac{1}{3}.
   \end{split}
   \label{eq:youngpoisson}
   \end{equation}

With these assumptions, we can construct expressions for the two principal
elastic contributions to the total energy arising from the presence of a
mismatched island on a substrate, viz.\ the energy due to the mutual strain
between the island and the substrate, and the island-island interaction
energy mediated by the substrate:
   \begin{equation}
   \Delta E_{\text{elastic}} = \Delta E_{\text{strain}} + E_{\text{interaction}}.
   \label{eq:straininter}
   \end{equation}

The elastic energy per unit volume (more precisely unit area in the present
case) of a strained uniform adsorbed layer is
   \begin{equation}
   \frac{d E^{\text{layer}}_{\text{elastic}}}{dV} = \frac{1}{2}E_A
   \alpha^2,
   \end{equation}
$E_A$ being the Young modulus of the adsorbate. Hence, a portion of width $d$
of this $\frac{\sqrt{3}}{2} \theta$-thick layer has an energy
   \begin{equation}
   E^{\text{layer}}_{\text{elastic}} = \frac{2}{\sqrt{3}} \mu_A \alpha^2
   \theta d.
   \end{equation}
Similarly, we can write
   \begin{equation}
   E^{\text{island}}_{\text{elastic}} = \frac{1}{2} E_A \alpha^2
   v(L,h),
   \end{equation}
where $v(L,h)$ is a function (to be determined) having units of volume.
CJB\cite{combe2001} write this as
   \begin{equation}
   v(L,h) = R(r)V,
   \label{eq:vredef}
   \end{equation}
where $r=h/L$ is the aspect ratio, $V$ is the volume of the island, and $R(r)$
is a dimensionless function bounded between 0 and 1. Evidently, for scaling
reasons, any $v(L,h)$ can be written in this form. In the present work,
however, we choose to determine $v(L,h)$ without invoking this scaling
ansatz. This choice is mainly dictated by accuracy considerations: in order
to fit $R(r)$ to numerical data, it is necessary to divide the computed
elastic energy by the volume, thereby reducing the relative importance of
large islands (see appendix \ref{app:2} for more details).

We find that an excellent fit to the Lennard-Jones data is obtained with
   \begin{equation}
   v(L,h) = \frac{\sqrt{3}}{2} \frac{L^2}{c} \left[ 1 - e^{-c h/(L-h)}
   \right],
   \label{eq:vlh}
   \end{equation}
where $c$ is the only (unitless) parameter to be adjusted to the data.
Numerical calculations on the LJ system yield $c=13.5$ (see
Appendix~\ref{app:1} for details). Ratsch and Zangwill\cite{ratsch1993}
have developed a similar function from theoretical considerations; the
only difference is the denominator of the exponent, where they use $L$
instead of $L-h$ and find $c = 2\sqrt{3} \pi \approx 10.9$.

In summary we have, for the elastic energy due to the strain between the
island and the substrate:
   \begin{equation}
   \Delta E_{\text{strain}} = \frac{2}{\sqrt{3}} \mu_A \alpha^2 \left[
   \frac{L^2}{c} \left( 1 - e^{-c h/(L-h)} \right) +zd -\theta d \right],
   \label{eq:estrain}
   \end{equation}
where we include the contribution arising from the presence of a wetting
layer of thickness $z$ in the system with islands.

The substrate-mediated interaction energy between two islands a distance $d$
apart --- the second term in Eq.\ \eqref{eq:straininter} --- can be written
as a surface integral,
   \begin{equation}
   E_{\text{isl-isl}}(d) = \int_{-\infty}^{\infty} u_x(x) f(x-d) dx,
   \end{equation}
where $u_x(x)$ is the displacement field of the first island and $f$ is the
surface force distribution of the second island. Using \eqref{eq:fdist} and
\eqref{eq:utilde}, we get:
   \begin{multline}
   E_{\text{isl-isl}}(d) = - \frac{4 \pi \mu_S \kappa^2}{p L^2}
   \left[ 3 L^4 + 2 d^2 L^2 + 4 d L^3 \log{\left| \frac{d-L}{d+L}
   \right|} \right. \\
   \left. + 2 (d^4 - 3 d^2 L^2) \log{\left| \frac{d^2- L^2}{d^2}
   \right|}\right].
   \end{multline}
Note that this expression remains finite when $d \to L$. Since we are
interested in the interaction energy for an array of islands, we have to sum
up all contributions coming from the islands at position $..., -2d, -d, d,
2d, ...$ In order to get a simple expression for our model, we may replace
$E_{\text{isl-isl}}$ by the first few terms of its asymptotic series,
   \begin{equation}
   E_{\text{isl-isl}} \approx \frac{2 \pi^2 \mu_S L^2 \kappa^2}{9} \left(
   \frac{2 L^2}{3 \pi d^2} + \frac{L^4}{5 \pi d^4} + \frac{3 L^6}{35 \pi
   d^6} + \dots \right).
   \end{equation}
The sum of all possible contributions can now be carried out exactly:
   \begin{multline}
   E_{\text{interaction}} = \sum_{\underset{j\neq0}{j=-\infty}}^{\infty}
   E_{\text{isl-isl}}(j \, d) \\
   \approx  \frac{4 \pi^3 \mu_S L^2 \kappa^2}{81} \left( \frac{L^2}{d^2} +
   \frac{\pi^2 L^4}{50 d^4} + \frac{\pi^4 L^6}{1225 d^6}
   \right).
   \end{multline}

We still have to determine the dependence of $\kappa$ [cf. Eq.
\eqref{eq:utilde}] on the parameters of our model. Let us first recall that
this quantity is proportional to the amplitude of the force distribution
exerted on the substrate by the island and, therefore, it also determines the
amplitude of the displacement field in the substrate. We assume $\kappa$ to
have the following form:
   \begin{equation}
   \kappa = \alpha \; \kappa_{\mu}(\mu_A,\mu_S) \; \kappa_{\text{geo}}(h,L),
   \end{equation}
where $\kappa_{\mu}$ depends only on the elastic coefficients of the
susbtrate and the adsorbate, and $\kappa_{\text{geo}}$ depends only on the
geometry of the island. $\kappa$ is linear in $\alpha$ because Eq.\
\eqref{eq:utilde} has itself been derived within the framework of the linear
theory of elasticity.

For $\kappa_\mu$ we propose the {\it ad hoc} expression
   \begin{equation}
   \kappa_{\mu} = \frac{\mu_A}{\mu_A+\mu_S}
   \end{equation}
which satisfies the three important limiting cases
   \begin{eqnarray}
   \mu_S \gg \mu_A & \Longrightarrow & \kappa_{\mu} \rightarrow 0 \nonumber
   \\
   \mu_S \ll \mu_A & \Longrightarrow & \kappa_{\mu} \rightarrow 1 \\
   \mu_S = \mu_A & \Longrightarrow & \kappa_{\mu} = \frac{1}{2}.
   \nonumber
   \end{eqnarray}
The first relation corresponds to the case of a substrate which is much more
rigid than the adsorbate, the second is the opposite case, and the third is
the case of equally rigid substrate and adsorbate. The geometry factor
$\kappa_{\text{geo}}$ can be determined by fitting to the LJ data. We find
the following expression to yield good results:
   \begin{equation}
   \kappa_{\text{geo}} = \frac{1}{4} \Big( 1 - e^{-a_1 L + a_2}
   \Big) \Big(1 - e^{-b_1 h/(L-h) + b_2}\Big).
   \label{eq:kapgeo}
   \end{equation}
The first factor, $\frac{1}{4}$, is such that $\partial_x u_x$, the
$x$-component of the substrate's strain tensor, is always between 0 and
$\alpha$. The second factor ensures that $\kappa_{\text{geo}}$ goes rapidly
to 1 when $L$ is larger than a few atoms. The last factor is a scale
invariant term which is maximum when the aspect ratio $\frac{h}{L}$ is 1.
After fitting the displacement field to the LJ data (see
Appendix~\ref{app:1}), we find the following set of values for the
parameters entering the above expression:
   \begin{equation}
   \begin{split}
   a_1 = 12.6/r_{\text{eq}}, \quad a_2 = 0.028, \\ b_1 = 0.033, \quad b_2
   = -1.35,
   \end{split}
   \end{equation}
where $r_{\text{eq}}$ is the equilibrium interatomic distance (see previous
section).

Putting everything together, we obtain the following long expression for the
interaction energy in Eq.\ \eqref{eq:straininter}:
   \begin{widetext}
   \begin{equation}
   E_{\text{interaction}} =  \frac{4 \pi^3 \mu_S L^2}{81} \left[
   \frac{\alpha}{4} \frac{\mu_A}{\mu_A + \mu_S}  \Big( 1 - e^{-a_1 L + a_2}
   \Big) \Big(1 - e^{-b_1 h/(L-h) + b_2}\Big) \right]^2 \left(
   \frac{L^2}{d^2} + \frac{\pi^2 L^4}{50 d^4} + \frac{\pi^4 L^6}{1225 d^6}
   \right).
   \label{eq:einteraction}
   \end{equation}
   \end{widetext}
We are now in position to construct the phase diagram of our model as a function
of the various control parameters.

\section{Results and Discussion \label{sec:results}}

\subsection{Phase Diagrams}

The energy difference $\Delta E$ between a system in which the adsorbate
atoms form islands and a system in which they form a uniform layer on top of
the substrate, Eq.\ \eqref{eq:deltae}, can be expressed as
   \begin{equation}
   \Delta E = \Delta E_{\text{surface}} + \Delta E_{\text{strain}} +
   E_{\text{interaction}},
   \end{equation}
where the three terms are given by Eqs.\ \eqref{eq:esurface},
\eqref{eq:estrain}, and \eqref{eq:einteraction}, respectively. We now proceed
to determine the equilibrium configurations of the system as a function of
the control parameters, viz.\ binding energies, mismatch and coverage. The
parameter space is sampled as follows:
   \begin{eqnarray*}
   \epsilon_{AA} & = & 0.7, 0.8, \dots, 1.2, 1.3 \\
   \epsilon_{SA} & = & 0.7, 0.8, \dots, 1.2, 1.3 \\
   \alpha & = & 0\%, 1\%, \dots,9\%, 10\% \\
   \theta & = & 1, \dots, 14, 15,
   \end{eqnarray*}
and $\epsilon_{SS} = 1$ sets the energy scale. We took positive values of
$\alpha$ only since $\Delta E$ depends quadratically on this parameter. For
every possible combination of the above parameters, we determine $h$, $L$,
$d$ and $z$ such that $\Delta E$ is minimal. If the minimum $\Delta E$ is
larger than zero, the equilibrium configuration is a uniform adsorbate layer
on the substrate; this phase is known as the ``Frank-Van der Merwe''
phase. If it is less than zero, the equilibrium phase depends on the
values of $h_{\text{min}}$, $L_{\text{min}}$, $d_{\text{min}}$, and
$z_{\text{min}}$ which minimize $\Delta E$. A few different situations can
arise (we omit the ``min'' subscript to facilitate reading):
   \begin{enumerate}
   \item If $d \to \infty$, the islands undergo ripening, either directly on
   the substrate ($z=0$) or on a wetting layer ($z \neq 0$);
   \item If $d$ is finite and $L < d$, the equilibrium configuration is an
   array of islands either on the substrate (``Volmer-Weber''
   phase\cite{volmer1926}) or on a wetting layer (``Stranski-Krastanov''
   phase\cite{stranski1937}).
   \item If $d$ is finite and $L = d$, the system is ``cracked'', that is,
   islands touch at their base.
   \end{enumerate}

\begin{figure}[!t]
\centering
\includegraphics[width=3in]{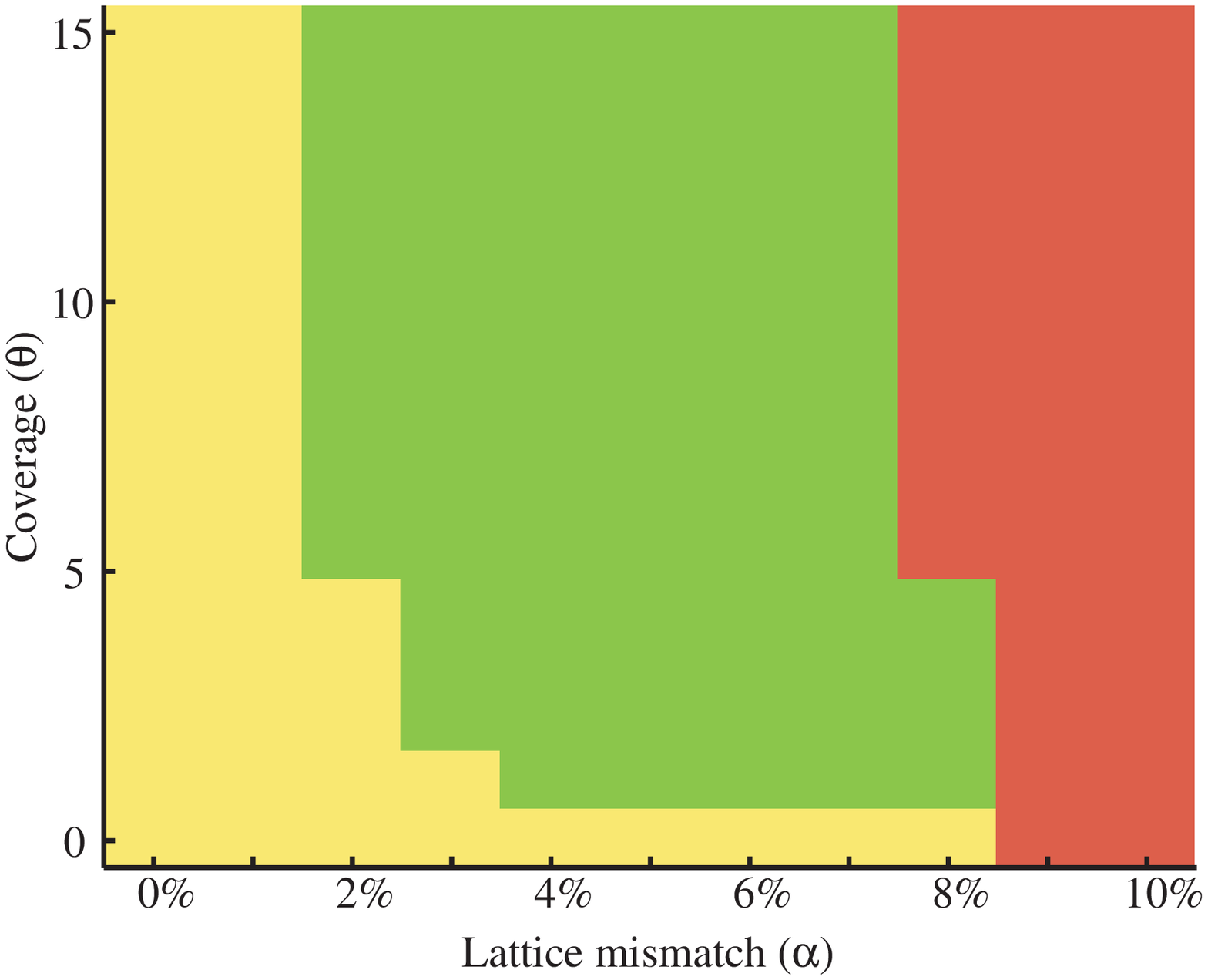}
\caption{
(Color online)
A typical phase diagram; here $\epsilon_{AA}=1$ and $\epsilon_{SA} = 1.1$.
The three phases are:
\protect\keybox{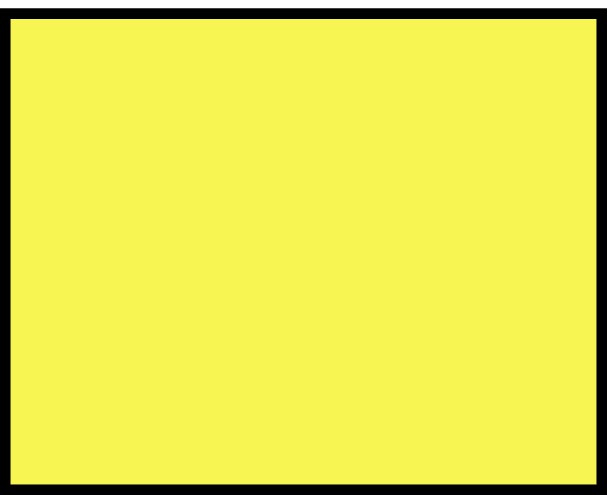} Frank-van der Merwe;
\protect\keybox{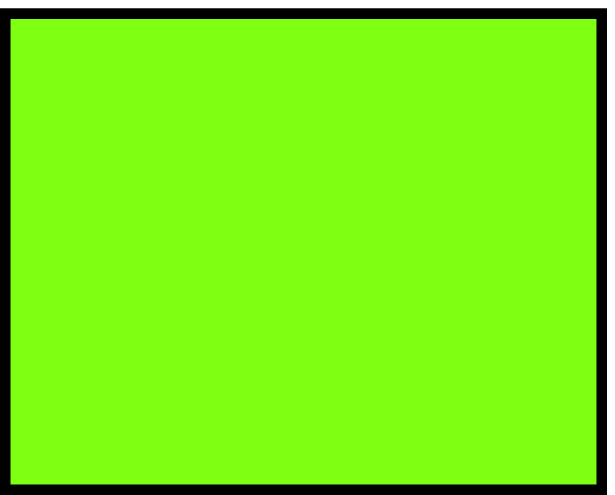} ripening islands with wetting layer;
\protect\keybox{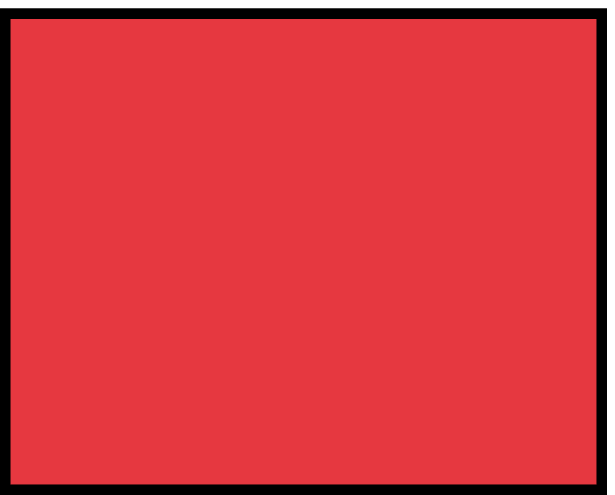} ripening islands without wetting layer.}
\label{fig:exdphase}
\end{figure}

\begin{figure}
\centering
\includegraphics[width=3in]{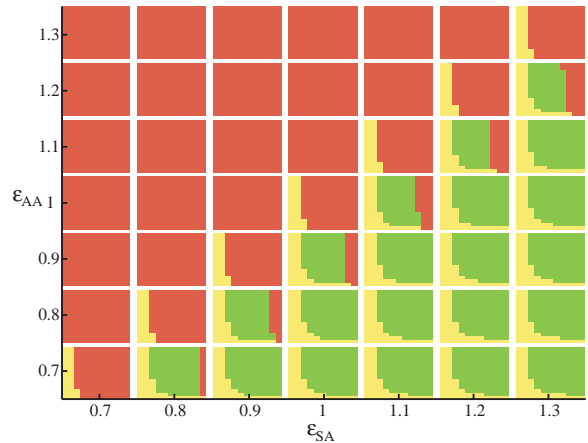}
\caption{
(Color online)
Phase diagrams for all values of the binding energies considered; here
$\epsilon_{SS}=1$ and $z_0=0.39$. Each small image is a phase diagram in the
$\alpha-\theta$ plane similar to that of Fig.\
\ref{fig:exdphase}. The colors correspond to the different phases:
\protect\keybox{fig/FMkey.eps} Frank-van der Merwe;
\protect\keybox{fig/R1key.eps} ripening islands with wetting layer;
\protect\keybox{fig/R2key.eps} ripening islands without wetting layer.}
\label{fig:dphase}
\end{figure}

Following the example of Daruka et al.\cite{daruka1997}, we produce a
set of phase diagrams in the $\alpha$-$\theta$ plane. Figure
\ref{fig:exdphase} shows a typical phase diagram for specific values of
the binding energies ($\epsilon_{AA}=1$ and $\epsilon_{SA}=1.1$). Such
plots are collected in Fig.\ \ref{fig:dphase} for the whole array of
values of $\epsilon_{AA}$ and $\epsilon_{SA}$ investigated.

We observe, first, that only one phase is possible when $\epsilon_{AA} >
\epsilon_{SA}$ (above the main diagonal of the plot), namely ripening islands
without wetting layer. This is not surprising since it corresponds to the
non-wetting case in which all terms of $\Delta E$ but
$E_{\text{interaction}}$ are negative. A more important conclusion that
can be drawn from Fig.\ \ref{fig:dphase} is that {\em no set of control
parameters leads to an array of islands as a stable configuration}. This
is in disagreement with the results of CJB.\cite{combe2001} The
discrepancy can be traced back to the choice of $z_0$, the
characteristic length for the decay of the adsorption energy at the
surface, discussed in Section \ref{subsubsec:surf}. The value of this
parameter (0.39) was obtained, in the present model, by fitting to the
LJ potential, while CJB chose it in an {\em ad hoc} manner.

\subsection{Extended model}

\begin{figure}[!ht]
\centering
\includegraphics[width=3in]{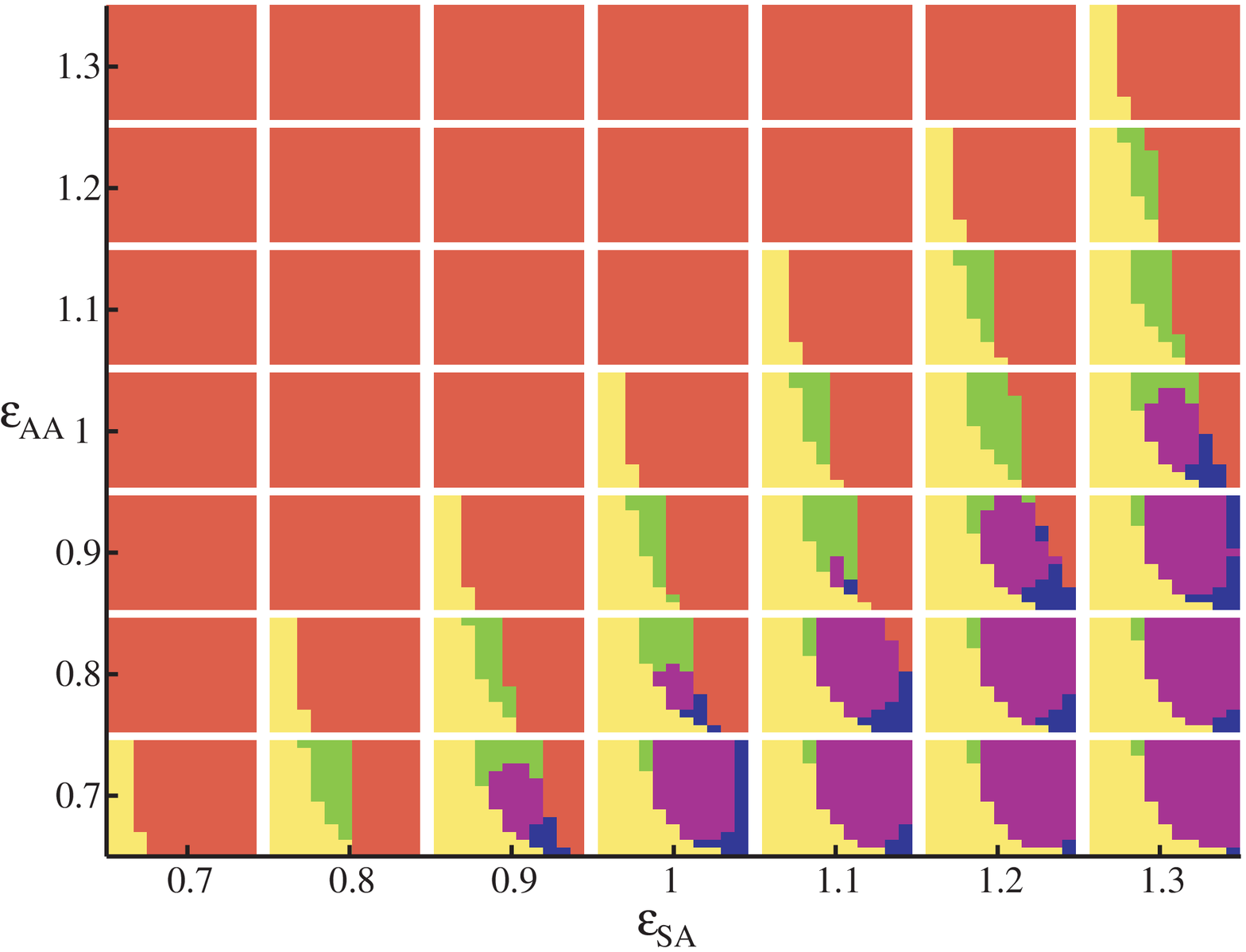}
\caption{
(Color online)
Same as Fig.\ \ref{fig:dphase}, but for $z_0=3.0$. The various
phases are as follows:
\protect\keybox{fig/FMkey.eps} Frank-van der Merwe;
\protect\keybox{fig/R1key.eps} ripening islands with wetting layer;
\protect\keybox{fig/R2key.eps} ripening islands without wetting layer;
\protect\keybox{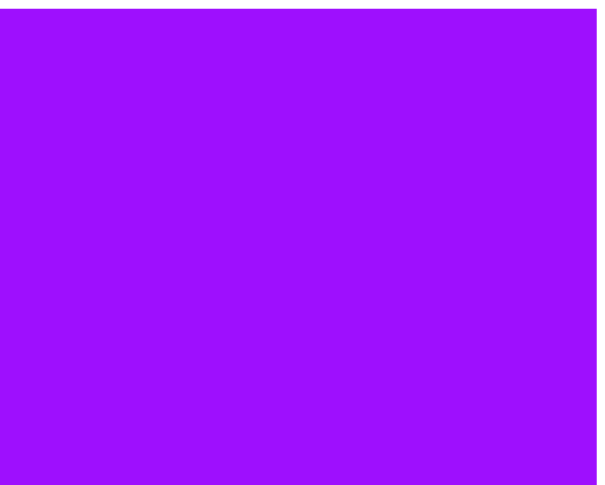}  cracks;
\protect\keybox{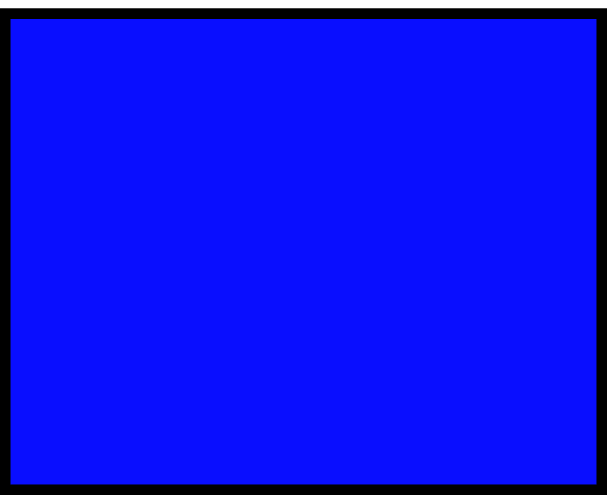} Volmer-Weber.}
\label{fig:dphase_3}
\end{figure}

We may relax the constraint on the value of $z_0$ for a moment and set $z_0 =
3$, close to the value used by CJB, resulting in an increase of the influence
of the substrate on the adatoms higher above the interface. The introduction
of this new length scale in the problem allows some new features to appear.
The corresponding phase diagrams are displayed in Fig.\ \ref{fig:dphase_3}.
Consistent with CJB, we now observe the existence of two different stable
phases. In addition to the stable array of islands, we now find an
equilibrium state consisting of islands touching at their base. Since this
morphology is somewhat better described by a flat layer with very narrow
troughs, we will refer to it as the ``cracks'' phase (shown in red in
Fig.~\ref{fig:dphase_3}). The Stranski-Krastanov phase never appeared in the
parameter space we considered.

It is not clear that potentials with $z_0=3$ do in fact exist, but it is
certainly conceivable, in the case for instance of semiconductors, that the
decay length of the surface energy is significantly larger than that of the
LJ potential and, as a consequence, stable phases would exist. Consideration
of this extended system is consistent with our long term objectives since
$z_0$ can be included as an adjustable parameter in the KMC calculations.

\begin{figure}[!ht]
\centering
\includegraphics[width=3in]{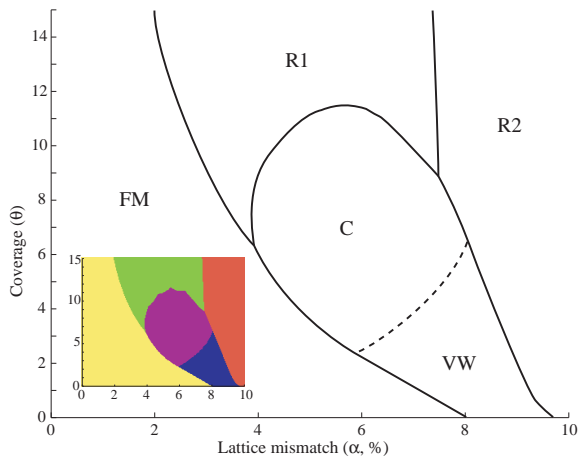}
\caption{
(Color online)
Phase diagram for the system with $z_0=3.0$, $\epsilon_{AA} = 1$, and
$\epsilon_{SA} = 1.3$. The various phases are: Frank-van der Merwe (FM),
ripening islands with wetting layer (R1), ripening islands without wetting
layer (R2), cracks (C), and Volmer-Weber (VW). The dashed line indicates the
smooth transition between the VW and the C phases. Inset: original phase
diagram used to construct the main graph.}
\label{fig:dphase_3a}
\end{figure}

\begin{figure}[!b]
\centering
\includegraphics[width=3in]{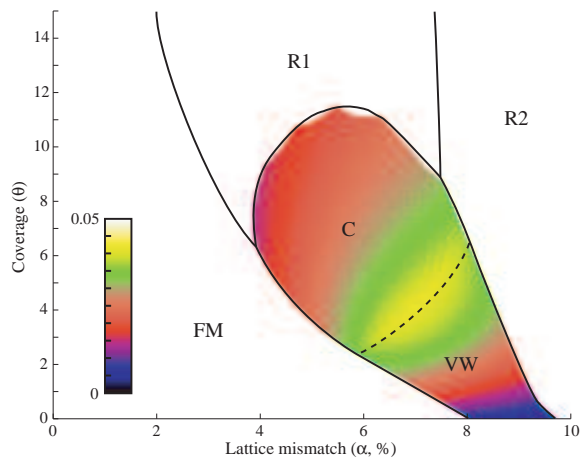}
\caption{
(Color online)
Island density $1/d$ in the two stable phases as a function of coverage
$\theta$ and lattice mismatch $\alpha$. Here, $z_0=3.0$, $\epsilon_{AA} = 1$,
and $\epsilon_{SA} = 1.3$.}
\label{fig:islanddensity}
\end{figure}

In order to get more detailed information about the stable phases, we
selected one of the phase diagrams of Fig. \ref{fig:dphase_3} ($\epsilon_{SA}
= 1.3$, $\epsilon_{AA} = 1$) and repeated the calculations on a finer grid
(200$\times$200). The resulting diagram is shown in Fig.\
\ref{fig:dphase_3a}. Note that, in this graph, the layout of the phase
boundaries have been drawn as a ``guide to the eye'' using the original data
(shown in the inset); we do not know the analytical form (if it exists) of
these curves. (The slightly ``zigzagging'' behaviour of the boundary is a
consequence of allowing $z$ to take only integer values in the minimization
process.) We find that the transition is not sharp between the Volmer-Weber
(VW) and ``cracks'' (C) phases (hence the dashed line).
Figure~\ref{fig:dphase_3a} has some features similar to the phase diagram
computed by Daruka et al.\cite{daruka1997}, in particular the shape of the
FM, R1 and R2 phases.

The Frank-van der Merwe (FM) and the two ripening (R1 and R2) phases have no
interesting intrinsic features; the first is flat, and the two others
correspond to the minimizing parameters going to infinity. We are therefore
more concerned about the two stable phases (C and VW). A characteristic
quantity often measured is the island density $n$ on the surface. In our
model, since the center-to-center distance between islands is $d$, we simply
have $n=1/d$. Figure~\ref{fig:islanddensity} shows how the density varies as
a function of coverage and lattice misfit for the binding energies selected.
In the C phase, this density is simply the cracks density, that is, the
number of cracks per unit length.

\begin{figure}[!hb]
\centering
\includegraphics[width=3in]{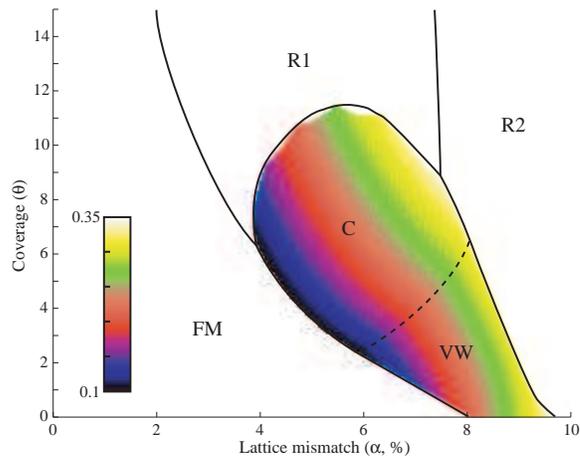}
\caption{
(Color online)
Aspect ratio $h/L$ in the two stable phases as a function of coverage
$\theta$ and lattice mismatch $\alpha$. Here $z_0=3.0$, $\epsilon_{AA} = 1$,
and $\epsilon_{SA} = 1.3$.}
\label{fig:aspectratio}
\end{figure}

We have also computed the aspect ratio of the islands as a function of
coverage and lattice mismatch; this is shown in Fig.~\ref{fig:aspectratio}.
The global behavior is as expected: as $\alpha$ increases, the elastic
relaxation process becomes more and more efficient and the islands can afford
an increase of their surface; this explanation holds for the cracks phase as
well.

In both Figs. \ref{fig:islanddensity} and \ref{fig:aspectratio}, the
quantities shown by the color scale are conspicuously continuous at the VW-C
boundary. This was to be expected since the very definition of these phases
implies no jump in any quantity (VW phase for $L<d$, C phase for $L=d$). This
boundary actually is the only second order phase transition; all others are
first order.

To our knowledge, a cracks phase has never been reported. We do not know if
an equivalent 3D phenomenon has been observed experimentally. In our model,
the occurence of such a phase is obviously related to the interaction energy
term. Have we had assumed that $E_{\text{interaction}}$ is infinite when the
islands touch (as did CJB\cite{combe2001}), the C phase would have been
entirely precluded. A stronger island-island repulsion might also lead to the
appearance of a Stranski-Krastanov (SK) phase for intermediate coverage.
While it remains to be verified, we expect this cracks phase to be very close,
at finite temperature for instance, to a SK phase; this is however beyond the
scope of the present work.

\section{Conclusion \label{sec:conclu}}

We have presented a simple analytical model for the determination of the
stable phases of strained heteroepitaxial systems in $(1+1)$ dimensions. The
model was developed with a view of carrying out KMC simulations of the
dynamics of the formation of islands. This is a very difficult task, but
already we have made progress in this direction, on which we will report in a
subsequent publication. In order for the present model to be exportable to a
KMC code, all expressions were adjusted to an atomistic Lennard-Jones system.
The present calculations reveal that, for parameters which are consistent
with the Lennard-Jones model, the array of islands is not a stable
configuration of the system. If full consistency of the parameters is not
imposed, and in particular if we relax the value of the decay length for the
adsorption energy ($z_0$), then a stable array of islands arises. We argue
that $z_0$ may in fact be viewed as an adjustable parameter, which can be
used to describe systems other than Lennard-Jones, in particular
semiconductors. Our calculations also reveal, in these conditions, the
formation of a cracks phase --- an array of islands touching at their base.
While much work remains to be done, the present model is a first step in our
aim of better understanding the formation of dislocation-free arrays of
islands.

\begin{acknowledgments}

One of us (PT) is grateful to P. Jensen and J.-L. Barrat, from the
``Laboratoire de physique de la mati\`ere condens\'ee et nanostructures,
Universit\'e Claude-Bernard Lyon-1'', for hospitality and fruitful
discussions. This work was supported by grants from the Natural Sciences and
Engineering Research Council (NSERC) of Canada and the ``Fonds qu\'eb\'ecois
de la recherche sur la nature et les technologies'' (FQRNT) of the Province
of Qu\'ebec. We are indebted to the ``R\'eseau qu\'eb\'ecois de calcul de
haute performance'' (RQCHP) for generous allocations of computer ressources.

\end{acknowledgments}

\appendix

\section{Computational Details \label{app:1}}

We present here some details of the method used to fit the free parameters
arising in the derivation of the continuous model presented in Section
\ref{sec:model}.

In what follows, the cutoff radius of the potential has been fixed to $r_c =
3.2$, i.e., midway between the 5th and 6th neighbour shells. The equilibrium
distance between substrate atoms is set to 1 so that $\sigma_{SS} = 1/1.1119
= 0.8993$ (cf.\ Table \ref{tab:1}). In all calculations, the substrate has
thickness between 50 and 100 layers, with the lower three maintained fixed to
mimic the presence of the bulk. For every configuration, the energy was
determined by relaxing the positions of the atoms using a conjugate-gradient
algorithm.

\subsection{Surface Energy}

\begin{figure}
\centering
\includegraphics[width=3in]{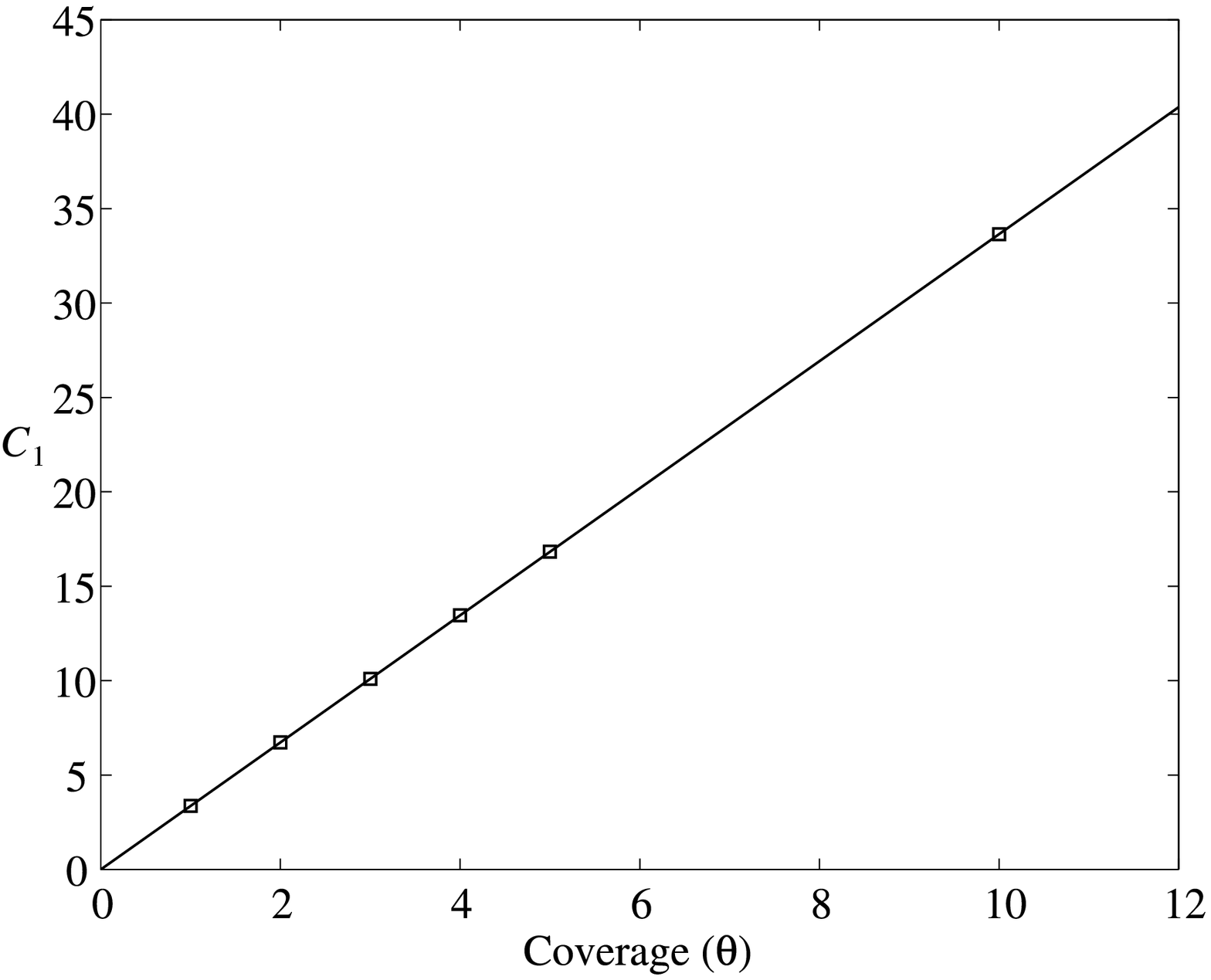}
\raisebox{5.6cm}{\makebox[0pt][r]{a) \hspace{6.5cm}}}
\includegraphics[width=3.05in]{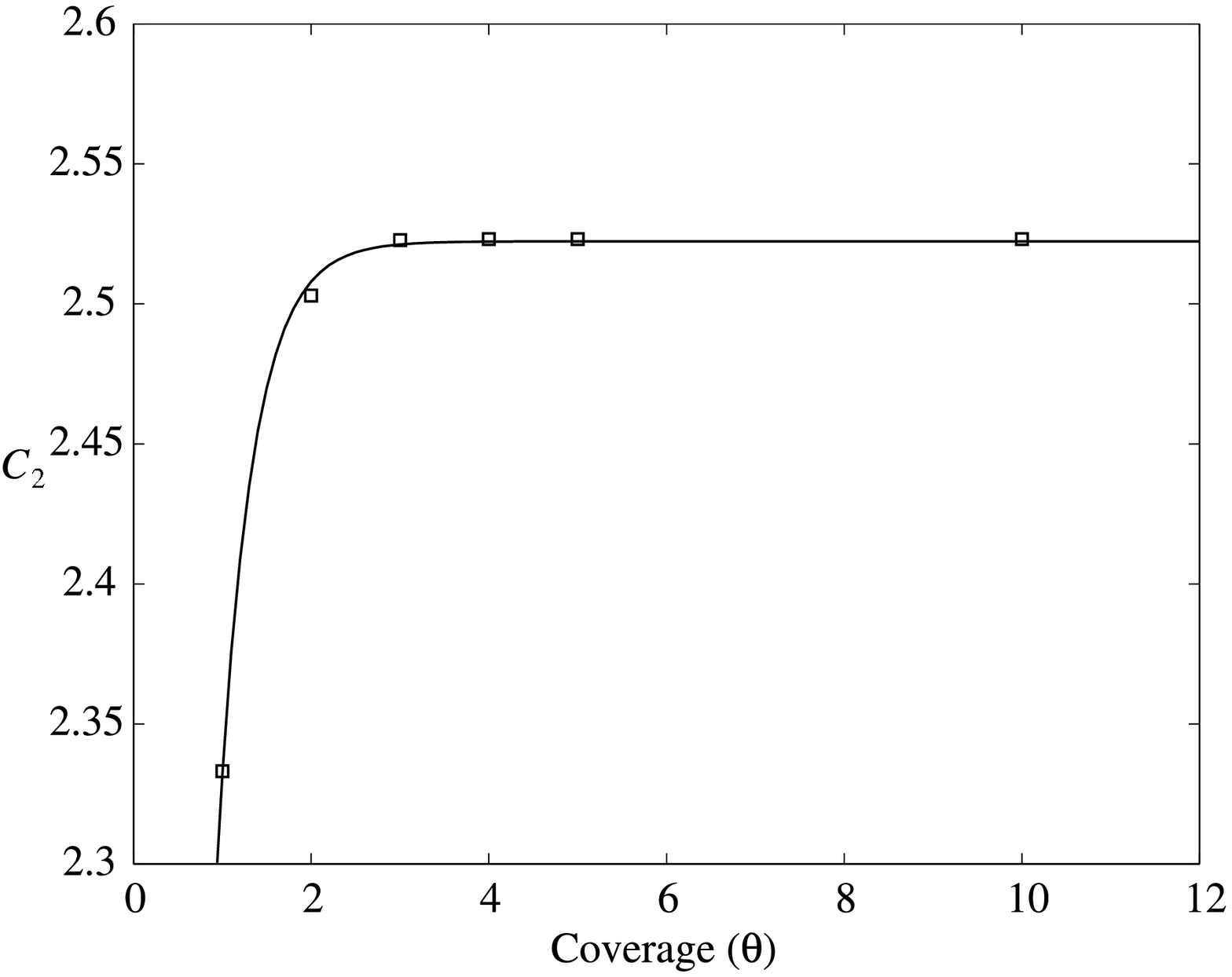}
\raisebox{5.6cm}{\makebox[0pt][r]{b) \hspace{6.5cm}}}
\caption{Comparison between the numerical values of $C_1$ and $C_2$ (in
equation~\ref{eq:c1c2}) and the theoretical curves (see text). (a) The curve has no
free parameter. (b) The curve is the best fit to the data.
}
\label{fig:coef12}
\end{figure}

The parameters $B$ and $z_0$ in Eq.\ \eqref{eq:enadsorpa} have been fitted to
systems composed of 50 substrate layers and coverage $\theta \in
\{1,2,3,4,5,10\}$. These configurations have been relaxed for a set of
combinations of energy parameters $(\epsilon_{AA}, \epsilon_{SA}) \in \{0.8,
0.9, 1, 1.1,1.2 \}^2$. Altogether, the minimum total energies of 150
different systems have been computed. For every system, the difference
between the total energy and the total energy of the equivalent system with
$\epsilon_{SA} = \epsilon_{AA} = \epsilon_{SS}$ has been calculated. The
numerical values for $E_{A}^{\text{layer}} - E_{S}^{\text{layer}}$ in
\eqref{eq:enadsorpa} have then been fitted to an equation of the form
   \begin{equation}
   E_{A}^{\text{layer}} - E_{S}^{\text{layer}} = C_1(\theta)
   (\epsilon_{AA}-\epsilon_{SS}) + C_2(\theta) (\epsilon_{AA}-
   \epsilon_{SA}).
   \label{eq:c1c2}
   \end{equation}
Figure \ref{fig:coef12} shows the dependence of $C_1$ and $C_2$ on coverage.
$C_1$ is completely determined (i.e., there are no free parameters), given by
$C_1 = d u_0 \theta$, where $u_0 = 3.364$ is the cohesive energy for a cutoff
radius of 3 (see Table~\ref{tab:1}). $C_2$ was fitted to a curve of the form
$B(1-e^{-\theta/z_0})$; the fit yields $B=2.53$ and $z_0 = 0.39$.

\subsection{Elastic Energy and Island-island Interaction Energy}

\begin{figure}[!t]
\centering
\includegraphics[width=3.1in]{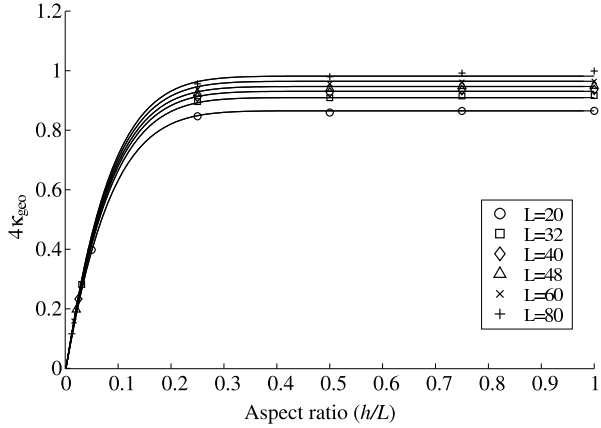}
\caption{Numerical data and theoretical curves for $\kappa_{\text{geo}}$
given by equation \eqref{eq:kapgeo}.}
\label{fig:kapgeo}
\end{figure}

The parameters $a_1$, $a_2$, $b_1$, $b_2$, and $c$ of Eqs.\ \eqref{eq:vlh} and
\eqref{eq:kapgeo} have been fitted to the same configurations as above. 30
different islands have been generated, of width $L$ in the range 20 to 80
(six values) and height $h$ in the range 1 to $L$ (five values). Each of
these configurations was relaxed with misfits of $+1 \%$ and $-1\%$. While
the relaxed positions of the atoms depend on the sign of $\alpha$, the energy
does not, as could be expected.

The numerical value of $\kappa_{\text{geo}}$ has been found using the
displacements of the atoms of the substrate; for every system, the amplitude
of the theoretical displacement field [which is close to, but not equal to
Eq. \eqref{eq:utilde} because of periodic boundary conditions] has been
adjusted to the displacements of the substrate atoms. Figure \ref{fig:kapgeo}
shows the agreement between the curves and the data.

\begin{figure}[!t]
\centering
\includegraphics[width=3.1in]{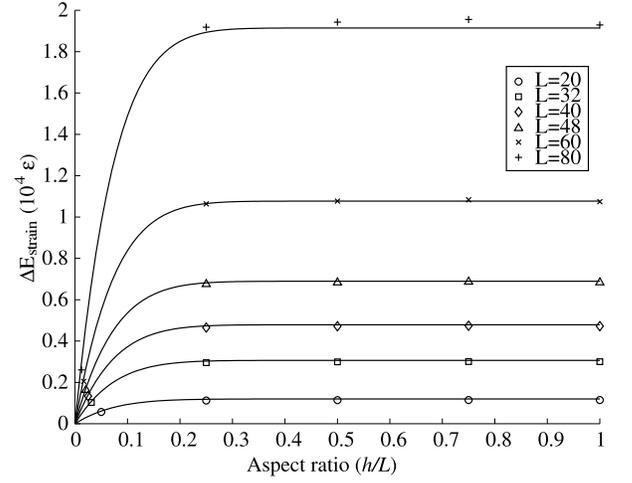}
\caption{Numerical data and theoretical curves for the strain energy given by
\eqref{eq:estrain}. }
\label{fig:eelast}
\end{figure}

Figure \ref{fig:eelast} shows the good agreement between the strain energy
obtained from the simulations and the analytical expression
\eqref{eq:estrain}. In Appendix~\ref{app:2} we elaborate on the choice
of this equation.

\section{Functional form of Eq. \lowercase{\protect\eqref{eq:estrain}}
\label{app:2}}

\begin{figure}[!ht]
\centering
\includegraphics[width=3.1in]{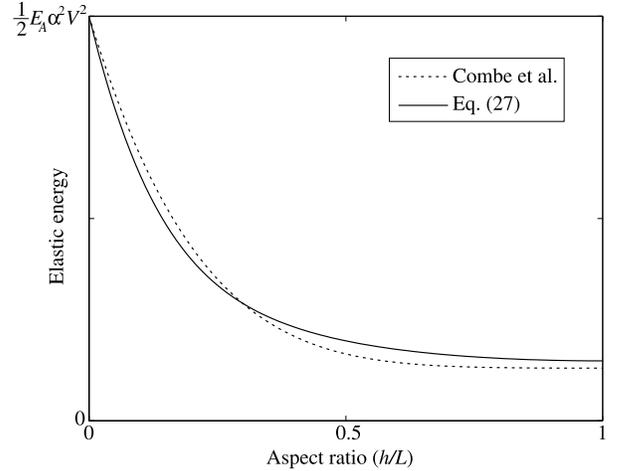}
\caption{Elastic energy for an island of given volume $V$ as a function
of its aspect ratio $h/L$.}
\label{fig:rr}
\end{figure}

We mentioned in Section \ref{subsubsec:elast} that a better fit to the
numerical data is obtained with a function of the form $E(L,h) = C v(L,h)$
[where $v(L,h)$ has units of volume], rather than $E(L,h) = C V R(h/L)$. We
show here a comparison between our expression for $v(L,H)$, Eq.\
\eqref{eq:vlh}, and that used by CJB\cite{combe2001}; the latter is obtained
by writing, first, the function $R(h/L)$ in terms of the quantities used in
the present paper:
   \begin{equation}
   R_{\text{CJB}}(r) = A + B e^{-C r/(1-r/2)},
   \label{eq:rcombe}
   \end{equation}
with $A=0.13$, $B=0.87$ and $C=-4.811$ and $r=h/L$. Since $v(L,h) = R(h/L)V$,
it is a simple matter to connect the two functional forms. We get
   \begin{equation}
   v_{\text{CJB}}(L,h) = \frac{\sqrt{3}}{2} h (L-h/2) \left( A + B
   e^{-C h/(L-h/2)} \right),
   \label{eq:vcombe}
   \end{equation}
which must be compared with Eq.\ \eqref{eq:vlh}, and
   \begin{equation}
   R(r) = \frac{\left( 1 - e^{-c r/(1-r)} \right)}{c r (1-r/2)}.
   \label{eq:rnous}
   \end{equation}
Figure \ref{fig:rr} shows the the elastic energy of an island of fixed volume
$V$ as a function of its aspect ratio, according to Eqs.\ \eqref{eq:rnous} and
\eqref{eq:rcombe}. The two curves are different, but their general behaviour
is very similar. Note in particular that the starting points coincide and
that the end points are less than 2\% apart.

\begin{figure}[!t]
\centering
\includegraphics[width=3.1in]{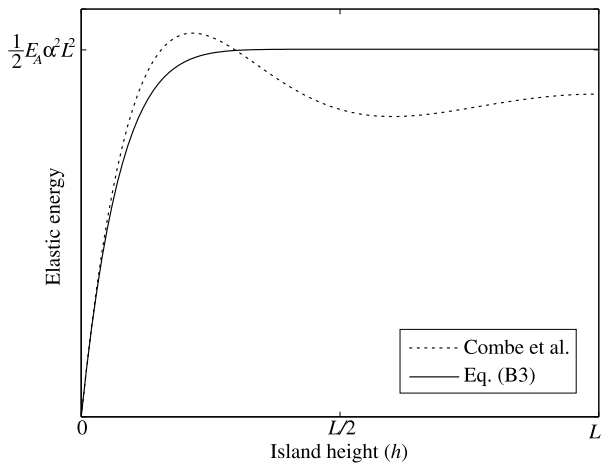}
\caption{Elastic energy for an island of given width $L$ as a function
of its aspect height $h$.}
\label{fig:vv}
\end{figure}

The situation is however very different for the elastic energy of an island
of fixed width as a function of height (Fig. \ref{fig:vv}). CJB's expression
for the energy yields a peak around $h/L = 0.2$ and has a minimum around
$h/L=0.6$; in contrast, our expression increases monotonically with height.
Our preference for the form of Eq. \eqref{eq:estrain} is mainly based on this
observation and the quality of the fit to the LJ data (see Fig.
\ref{fig:eelast}).

\end{document}